\documentclass[a4paper,11pt]{article}
\usepackage{amsmath,amssymb,theorem}
\usepackage[cm]{fullpage}
\usepackage[shortcuts]{extdash}
\usepackage{hyperref}
\usepackage[nosort]{cite}

\let\frac\undefined

\allowdisplaybreaks

\numberwithin{equation}{section}

\def\Maketitle{{\def\newpage{}\maketitle}}

\def\eq#1$$#2$${\begin{equation#1}#2\end{equation#1}}
\long\def\subeq#1{\begin{subequations}#1\end{subequations}}
\def\Split$$#1$${\begin{split}#1\end{split}}
\def\Align#1$$#2$${\begin{align#1}#2\end{align#1}}
\def\AlignAt#1$$#2$${\begin{alignat}{#1}#2\end{alignat}}
\def\Aligned#1{\begin{aligned}{}#1\end{aligned}}

\def\Gather#1$$#2$${\begin{gather#1}#2\end{gather#1}}
\def\Gathered#1{\begin{gathered}{}#1\end{gathered}}

\def\Multline#1$$#2$${\begin{multline#1}#2\end{multline#1}}
\def\Matrix#1{\begin{matrix}#1\end{matrix}}

\def\Cases#1{\begin{cases}#1\end{cases}}

\def\Ker{\mathop{\rm Ker}\nolimits}
\def\Tr{\mathop{\rm Tr}\nolimits}
\def\Re{\mathop{\rm Re}\nolimits}
\def\Im{\mathop{\rm Im}\nolimits}
\def\sign{\mathop{\rm sign}\nolimits}
\def\Res{\mathop{\rm Res}\limits}

\def\cA{{\mathcal A}}
\def\cC{{\mathcal C}}
\def\cF{{\mathcal F}}

\def\cH{{\mathcal H}}

\def\cN{{\mathcal N}}
\def\cO{{\mathcal O}}

\def\cT{{\mathcal T}}

\def\ve{\varepsilon}

\def\th{\mathop{\rm th}\nolimits}

\def\lcolon{\mathopen{\,:}}
\def\rcolon{\mathclose{:\,}}

\def\R{{\mathbb R}}
\def\bbT{{\mathbb T}}
\def\Z{{\mathbb Z}}

\def\cP{{\mathcal P}}
\def\cQ{{\mathcal Q}}

\def\e{{\rm e}}
\def\i{{\rm i}}

\def\lvac{\langle\textrm{vac}|}

\def\RSOS{\textrm{RSOS}}

\def\lbbrack{\mathopen{[\![}}
\def\rbbrack{\mathclose{]\!]}}

\makeatletter

\def\section{\@startsection{section}{1}{\z@}%
                                   {-3.5ex \@plus -1ex \@minus -.2ex}%
                                   {2.3ex \@plus.2ex}%
                                   {\normalfont\normalsize\bfseries}}
\def\subsection{\@startsection{subsection}{2}{\z@}%
                                     {-3.25ex\@plus -1ex \@minus -.2ex}%
                                     {1.5ex \@plus .2ex}%
                                     {\normalfont\normalsize\bfseries\itshape}}
\def\@seccntformat#1{\csname the#1\endcsname.~~}
\long\def\@makecaption#1#2{%
  \vskip\abovecaptionskip
  \sbox\@tempboxa{\small#1. #2}%
  \ifdim \wd\@tempboxa >0.9\hsize
  {\leftskip=0.05\hsize\rightskip=0.05\hsize\relax\small
    #1. #2\par}
  \else
    \global \@minipagefalse
    \hb@xt@\hsize{\hfil\box\@tempboxa\hfil}%
  \fi
  \vskip\belowcaptionskip}
\def\Appendix{\appendix
  \def\@seccntformat##1{Appendix~\csname the##1\endcsname.~~}}

\let\over\@@over
\let\atop\@@atop
\let\above\@@above
\let\overwithdelims\@@overwithdelims
\let\atopwithdelims\@@atopwithdelims
\let\abovewithdelims\@@abovewithdelims

\makeatother

\long\def\?#1{{\par\medskip\hrule\smallskip\noindent
{\bf What is missing:} #1\smallskip\hrule\medskip\par}}

\begin{document}

\title{Lattice models, deformed Virasoro algebra and reduction equation}
\author{Michael Lashkevich$^{a,b}$, Yaroslav Pugai$^{a,b}$, Jun'ichi Shiraishi$^c$ and Yohei Tutiya$^d$\\
	\parbox[t]{.8\textwidth}{\normalsize\it\raggedright
	\begin{itemize}\itemsep=\smallskipamount
		\item[$^a$]Landau Institute for Theoretical Physics, 142432 Chernogolovka, Russia
		\item[$^b$]Kharkevich Institute for Information Transmission Problems, 19 Bolshoy Karetny per., 127994 Moscow, Russia
		\item[$^c$]Graduate School of Mathematical Sciences, University of Tokyo, Komaba, Tokyo 153--8914, Japan
		\item[$^d$]Center for Basic Education and Integrated Learning, Kanagawa Institute of Technology, Atsugi, Shimo\-/Ogino 1030, Kanagawa 243--0292, Japan
	\end{itemize}
		ML:~lashkevi@landau.ac.ru, YP:~slava@itp.ac.ru, JS:~shiraish@ms.u-tokyo.ac.jp, YT:~tutiya@gen.kanagawa-it.ac.jp}
}

\date{}
\Maketitle

\begin{abstract}
We study the fused currents of the deformed Virasoro algebra (DVA). By constructing a homotopy operator we show that for special values of the parameter of the algebra fused currents pairwise coincide on the cohomologies of the Felder resolution. Within the algebraic approach to lattice models these currents are known to describe neutral excitations of the solid\-/on\-/solid (SOS) models in the transfer\-/matrix picture. It allows us to prove the closeness of the system of excitations for a special nonunitary series of restricted SOS (RSOS) models. Though the results of the algebraic approach to lattice models were consistent with the results of other methods, the lack of such proof had been an essential gap in its construction.
\end{abstract}

\section{Introduction}

The restricted solid\-/on\-/solid (RSOS) models~\cite{Andrews:1984af,Forrester:1985vsv} is a well\-/known class of two-dimensional integrable lattice models of classical statistical mechanics. Depending on the parameter they have several regimes, which have different scaling limits in the vicinity of the critical point. We will consider the so called regime~III, which provides minimal conformal models~\cite{Belavin:1984vu} at the critical point. The minimal conformal models $M(p,q)$ are labeled by a pair of coprime integers $0<p<q$, so that their central charges are
\eq$$
c=1-{6(q-p)^2\over pq}.
\label{cpq-def}
$$
It is natural to label the RSOS models by the same pair of integers and call them $\RSOS(p,q)$. For $q=p+1$ the minimal conformal models are unitary, which corresponds to the RSOS models with positive Boltzmann weights. For $q>p+1$ the minimal models are non\-/unitary, while the corresponding RSOS models contain negative Boltzmann weights. The models also depend on the temperature\-/like parameter $\epsilon\ge0$. The value $\epsilon=0$ corresponds to the critical point. For $\epsilon=\infty$ all Boltzmann weights except the largest in absolute value ones vanish, so that only a few configurations are allowed and have equal weights. This can be interpreted as the zero temperature limit.

Off the critical point correlation functions in the RSOS models were found~\cite{Lukyanov:1994re,Lukyanov:1996qs} in the framework of the algebraic approach~\cite{Jimbo:1992br,Davies:1992sva,Foda:1993fg,Jimbo:1994qp}. This approach makes it possible to calculate lattice correlation functions and, more generally, form factors of lattice operators. Form factors are matrix elements of operators defined in terms of the transfer\-/matrix approach with respect to eigenvectors of the transfer matrix. In the scaling limit in the vicinity of critical points the lattice form factors become well-known form factors of local operators of quantum field theory~\cite{Smirnov:1992vz}.

The most natural way to obtain form factors is the free field realization~\cite{Lukyanov:1993pn,Lukyanov:1994re,Lukyanov:1996qs}, and though it has been proving its validity for years, there are some open questions about it. We address the following problem. The free field representation is well defined for the unrestricted SOS models, but its consistency with the restriction is established \emph{a posteriori}, by checking the necessary properties for particular form factors. The SOS models contain two types of excitations: kinks and neutral `breathers'. Consistency of the free field representation in the kink sector is guaranteed by the Felder resolution~\cite{Felder:1988zp,Lukyanov:1996qs,Jimbo:1996vu}. The situation in the neutral sector seems to be more subtle. The closeness of the breather subsystem was never studied in the framework of the free field approach.

In this paper we consider the simplest example of the series $\RSOS(2,2s+1)$ ($s=2,3,\ldots$), where the restriction eliminates all kink excitations. On the infinite lattice every eigenvector $|A_{n_1}(\theta_1)\ldots A_{n_N}(\theta_N)\rangle$ of the transfer\-/matrix is labeled by a set of breather elementary excitations (or particles)~$A_n(\theta)$. Here the integer $n=1,\ldots,2s-2$ labels the type of a neutral particle, while the real number (`rapidity') $\theta$ modulo $\pi^2/\epsilon$ is in a one-to-one correspondence with the quasimomentum of the particle. In the scaling limit $\epsilon\to0$ the quantity $\th\theta$ is nothing but the velocity of the particle.

Assume the vacuum (the largest) eigenvalue of the transfer matrix to be equal~1. Then the excited state eigenvalue of the transfer matrix with the spectral parameter $u$ is given by a product of one-particle contributions:
\eq$$
t(u;\theta_1,\ldots,\theta_N)=\prod^N_{i=1}t^{(n_i)}\left({\theta_i\over\i\pi}-u\right),
\label{bbT(u)-def}
$$
where
\eq$$
t^{(n)}(v)
={\vartheta_4\left({v\over2}+{n\xi\over4};{\i\pi\over2\epsilon}\right)\vartheta_3\left({v\over2}-{n\xi\over4};{\i\pi\over2\epsilon}\right)
  \over\vartheta_3\left({v\over2}+{n\xi\over4};{\i\pi\over2\epsilon}\right)\vartheta_4\left({v\over2}-{n\xi\over4};{\i\pi\over2\epsilon}\right)}.
\label{bbT-nv}
$$
Here $\xi=2/(2s-1)$ and $\vartheta_i(z;\tau)$ ($i=1,\ldots,4$) denotes the Jacobi theta functions with the quasiperiods $1$ and~$\tau$ ($\Im\tau>0$). It is easy to check that the spectra of the particles labeled by $n$ and $n^*=2s-1-n$ coincide: $t^{(n)}(v)=t^{(n^*)}(v)$. In fact, in the restricted model the very particles are the same:
\eq$$
A_n(\theta)=A_{n^*}(\theta),
\qquad
n^*=2s-1-n,
\label{An-id}
$$
which means that
\eq$$
\lvac\cO|A_{n_1}(\theta_1)A_{n_2}(\theta_2)\ldots A_{n_N}(\theta_N)\rangle=\lvac\cO|A_{n^*_1}(\theta_1)A_{n_2}(\theta_2)\ldots A_{n_N}(\theta_N)\rangle
\label{matel-id}
$$
for any operator~$\cO$ consistent with the restriction. The free field representation provides explicit expressions for these matrix elements. Nevertheless, the left and right hand sides of~(\ref{matel-id}) are realized by expressions that look differently. In practical calculations of the one\-/particle form factors for the simplest operators $\cO$ the identity (\ref{matel-id}) was checked, but there were no general proof. The calculation of these matrix elements involves traces over the cohomologies of the Felder resolution, and the proof should involve homological argument. In the present paper we give such proof.

A similar problem appears for form factors of local operators in integrable perturbations of minimal conformal models. In the sine\-/Gordon model for special values of the coupling constant $\beta^2=2\beta_{BKT}^2/(2s+1)$ (where $\beta_{BKT}$ is the value corresponding to the Berezinskii\--Kosterlitz\--Thouless transition point) the $n$th and $n^*$th breathers have the same mass and $S$ matrix, as it was noticed by F.~Smirnov\cite{Smirnov:1990vm}. It was shown in~\cite{Smirnov:1990vm} that the identification of these particles amounts to the reduction condition, which turns the form factors of the sine\-/Gordon model into those of the perturbed minimal conformal model~$M(2,2s+1)$. In~\cite{Lashkevich:2014qna} we explained how to solve this condition in terms of the algebraic approach to breather form factors~\cite{Feigin:2008hs,Lashkevich:2013yja}. The solution was based on the usage of the algebra constructed in~\cite{Feigin:2007arXiv0705.0427}, which is a generalization of the deformed Virasoro algebra~\cite{Shiraishi:1995rp}. Here we use this construction as a guideline for our study of lattice models.

The structure of the paper is as follows. In section~\ref{sec-DVA} we recall the main facts about the deformed Virasoro algebra and its relation to the RSOS models. In section~\ref{sec-freefield} we describe the free field representation for the deformed Virasoro algebra and define the main objects: fused currents. Section~\ref{sec-tau-operator} is the central one. There we introduce the homotopy operator that relates the operators that represent the particles $A_n$ and $A_{n^*}$ and thus prove the identity~(\ref{matel-id}). In section~\ref{sec-fflimit} we consider the limit $\epsilon\to\i\pi/2$ that corresponds to the algebraic construction of form factors in the perturbed $M(2,2s+1)$ field theory. We show that this limit differs from the construction of~\cite{Lashkevich:2014qna} and explain the sense of this difference.

\section{Deformed Virasoro algebra}
\label{sec-DVA}

\subsection{Definition}

The deformed Virasoro algebra (DVA) \cite{Shiraishi:1995rp} is a family of algebras generated by the elements $T_m$ ($m\in\Z$) with a set of quadratic relations. It is parameterized by two complex numbers $\xi$ ($\Re \xi>0$) and $x$ ($0<|x|<1$).%
\footnote{In the notation of \cite{Shiraishi:1995rp} $q=x^{2\xi+2}$, $t=x^{2\xi}$.}
It is convenient to write the relations in terms of the currents $T(z)=\sum_{m\in\Z}T_mz^{-m}$:
\eq$$
\chi\left(z\over w\right)T(w)T(z)-\chi\left(w\over z\right)T(z)T(w)
={[\xi][\xi+1]\over[1]}\left(\delta\left(z\over wx^2\right) -\delta\left(z x^2\over w\right)\right),
\label{DVA-def}$$
where $\delta(z)=\sum_{m\in\Z}z^m$ and
\eq$$
\chi(z)={1\over1-z}{(x^{2\xi+2}z;x^4)_\infty(x^{-2\xi}z;x^4)_\infty\over(x^{2\xi+4}z;x^4)_\infty(x^{-2\xi+2}z;x^4)_\infty}.
\label{chi-def}
$$
Here we used the notation
\eq$$
[a]=x^a-x^{-a},
\qquad
(z;q_1,\ldots q_n)_\infty=\prod_{k_1,\ldots k_n\ge0}(1-zq_1^{k_1}\cdots q_n^{k_n}).
\label{main-notation}
$$
The deformation parameter $x$ is inside the unit circle: $|x|<1$. Below we will also use the notation
\eq$$
\Aligned{
\lbbrack v\rbbrack
&=x^{v^2/(\xi+1)-v}(x^{2v};x^{2\xi+2})_\infty(x^{2\xi+2-2v};x^{2\xi+2})_\infty(x^{2\xi+2};x^{2\xi+2})_\infty,
\\
\lbbrack v\rbbrack'
&=x^{v^2/\xi-v}(x^{2v};x^{2\xi})_\infty(x^{2\xi-2v};x^{2\xi})_\infty(x^{2\xi};x^{2\xi})_\infty.
}\label{bracket-function-def}
$$

The function $\chi(z)$ has a zero at the point $z=x^{2\xi}$. Hence, the product $T(z')T(z)$ admits a pole at $z'/z=x^{-2\xi}$, which, in fact, corresponds to the dynamic pole of form factors (see below). Its residue will be denoted as $T^{(2)}(z)$. The product $T^{(2)}(z')T(z)$ admits, in turn, a pole at $z'/z=x^{-3\xi}$ with the residue~$T^{(3)}(z)$ etc. More explicitly, let
\eq$$
T^{(n+1)}(z)
=\gamma_n\Res_{w=z}\left(T(x^{-\xi n}w)T^{(n)}(x^{\xi}z)\right){dw\over w},
\qquad
T^{(1)}(z)\equiv T(z).
\label{Fusion-DVA}
$$
It is convenient for us to take
\eq$$
\gamma_n={(x^4;x^4)_\infty(x^{4+2\xi};x^4)_\infty(x^{2+2(n+1)\xi};x^4)_\infty(x^{2+2n\xi};x^4)_\infty
  \over(x^2;x^4)_\infty(x^{2+2\xi};x^4)_\infty(x^{4+2(n+1)\xi};x^4)_\infty(x^{2n\xi};x^4)_\infty}.
\label{gamman-def}
$$
In the next subsection we will explain that the current $T^{(n)}(z)$ corresponds the $n$th neutral particle in the spectrum.

It was argued~\cite{Lukyanov:1994re,Lukyanov:1996qs} that the DVA can be treated as a dynamical symmetry algebra in the off\-/critical $\RSOS(p,q)$ model, if
\eq$$
\xi={p\over q-p},
\qquad
x=\e^{-\epsilon}.
\label{rx-RSOS}
$$
The spaces spanned on configurations on a half line, on which the corner transfer matrix acts in the thermodynamic limit, are identified with the irreducible representations $\cH_{kl}$, $1\le k\le p-1$, $1\le l\le q-1$ of the deformed Virasoro algebra. All physical operators can be expressed in terms of the vertex operators that act on these spaces.

\subsection{DVA and RSOS models}

In the framework of the RSOS models, an important role is played by the so called type~I vertex operators $\Phi(u)^{l'}_l$ ($l,l'\in\Z$, $1\le l,l'\le q-1$, $|l'-l|=1$). From the mathematical point of view these operators are primary $(1,2)$ operators of the DVA~\cite{Kadeishvili:1996ik,Awata:1996xt}, while physically they are half transfer matrices, necessary to define density matrices in the transfer matrix approach~\cite{Davies:1992sva,Foda:1993fg}. Here we skip the definitions and only cite their main properties. First of all, there is a commutation relation~\cite{Lukyanov:1994re,Lukyanov:1996qs}
\eq$$
\Phi(u_1)^{l'}_{\lambda'}\Phi(u_2)^{\lambda'}_l
=W\biggl[\Matrix{l'&\lambda'\\\lambda&l}\biggm|u_1-u_2\biggr]\Phi(u_2)^{l'}_\lambda\Phi(u_1)^\lambda_l,
\label{PhiPhi-commut}
$$
where $W[\cdots]$ are properly normalized Boltzmann weights of the RSOS model. Second, their commutation relation with the Virasoro algebra reads~\cite{Pugai:2002sft}
\eq$$
T^{(n)}(-x^{2v})\Phi(u)^{l'}_l
=t^{(n)}(v-u)\Phi(u)^{l'}_lT^{(n)}(-x^{2v}),
\label{TPhi-commut}
$$
where the function $t^{(n)}(u)$ is defined in~(\ref{bbT(u)-def}). Besides, there exists a grading operator $H$, which is defined as
\eq$$
\Phi(u)^{l'}_lx^{\alpha H}=x^{\alpha H}\Phi(u-\alpha/2)^{l'}_l,
\qquad
T^{(n)}(z)x^{\alpha H}=x^{\alpha H}T^{(n)}(x^{-\alpha}z).
\label{TH-commut}
$$
The product $T^{(n)}(z')T^{(n)}(z)$ has a pole at $z'=x^{-2}z$:
\eq$$
T^{(n)}(-x^{2u'})T^{(n)}(-x^{2u})
={1\over\pi c_n^2(u'-u+1)}+O(1),
\label{kinpole}
$$
where
\eq$$
c_n^2={2\epsilon\over\pi}(1-x^{2n\xi})^{-1}
\prod^n_{j=1}{(x^{2j\xi};x^4)_\infty(x^{4+2(j-1)\xi};x^4)_\infty(x^{2-2j\xi};x^4)_\infty(x^{2-2(j-1)\xi};x^4)_\infty
  \over(x^{-2j\xi};x^4)_\infty(x^{4-2(j-1)\xi};x^4)_\infty(x^{2+2j\xi};x^4)_\infty(x^{2+2(j-1)\xi};x^4)_\infty}.
\label{cn2-def}
$$
This pole provides the kinematic pole in the form factors defined below. Notice that the sign of $c_n^2$ can be negative as well as positive for real $\xi$ and~$x$. We will need their square roots $c_n$, whose signs will be chosen a little later. Define a `physically normalized' version of the DVA generator:
\eq$$
\bbT^{(n)}(u)=c_n T^{(n)}(-x^{2u}).
\label{bbT-def}
$$
From (\ref{kinpole}) we see that this operator satisfies the kinematic pole equation in the standard form, which provides the correct normalization of form factors.

Consider the products
\eq$$
E_{l\lambda l'\lambda'}=\Phi^*(u_M)^l_{\lambda'_{M-1}}\cdots\Phi^*(u_2)^{\lambda'_2}_{\lambda'_1}\Phi^*(u_1)^{\lambda'_1}_{l'}
\Phi(u_1)^{l'}_{\lambda_1}\Phi(u_2)^{\lambda_1}_{\lambda_2}\cdots\Phi(u_M)^{\lambda_{M-1}}_l,
\label{E-operator}
$$
where $\Phi^*(u)^{l'}_l=(-1)^{l'}(l'-l)\lbbrack l\rbbrack\Phi(u-1)^{l'}_l$ is a kind of `inverse' vertex operator:
\eq$$
\sum_{l'}\Phi^*(u)^l_{l'}\Phi(u)^{l'}_l=1.
\label{Phi*-Phi-rel}
$$
Each operator $E_{l\lambda l'\lambda'}$ in the corner transfer matrix space defines a lattice operator (a density matrix) $\cO_{ll'\lambda\lambda'}$ in the transfer\-/matrix picture (see~\cite{Jimbo:1994qp,Lukyanov:1996qs} for details). All other operators can be expressed as linear combinations of density matrices. The matrix elements with neutral excitations are related to the traces of the $E$ operators over irreducible representations $\cH_{kl}$ of the DVA according to
\eq$$
{}_k\lvac\cO_{l\lambda l'\lambda'}|A_{n_1}(\theta_1)\ldots A_{n_N}(\theta_N)\rangle_k
={1\over\cN_{kl}}
\Tr_{\cH_{kl}}\left(\lbbrack l\rbbrack\,x^{4H}E_{l\lambda l'\lambda'}\bbT^{(n_N)}(v_N)\cdots\bbT^{(n_1)}(v_1)\right),
\label{E-tr-rel}
$$
where $v_i=\theta_i/\i\pi$ and
\eq$$
\cN_{kl}=\sum^{q-1}_{\substack{l'=1\\l'-l\in2\Z}}\Tr_{\cH_{kl'}}\left(\lbbrack l'\rbbrack\,x^{4H}\right).
\label{chi-k-def}
$$
The subscript $k$ ($1\le k\le p-1$) enumerates vacuums in the regime~III RSOS model. There are $2p-2$ vacuums, one for each value of the parameter $k$ and the quantity $\nu=k-l\pmod2$. Recall that the vacuum number $k$ corresponds to the condition on infinity with staggered values $l_\infty$, $l_\infty+1$ of local heights such that~\cite{Forrester:1985vsv}:
\eq$$
0<qk-pl_\infty<p,
\qquad
0<l_\infty<q-1.
\label{kvac-def}
$$
The vacuums for $\nu=0$ and $1$ differ by a translation on one lattice step.

Thus the current $\bbT^{(n)}(v)$ plays the role of an operator creating the neutral excitation~$A_n(\i\pi v)$. The commutation relation (\ref{DVA-def}), in fact, defines a kind of Zamolodchikov\--Faddeev algebra
$$
S^{(m,n)}(v_1-v_2)\bbT^{(m)}(v_1)\bbT^{(n)}(v_2)=\bbT^{(n)}(v_2)\bbT^{(m)}(v_1)
$$
for generic values of $v_1-v_2$. The functions
\eq$$
S^{(m,n)}(v)=\prod^m_{i=1}\prod^n_{j=1}S(v+\xi(i-j)),
\qquad
S(v)={\chi(x^{-2v})\over\chi(x^{2v})}
={\theta_1\left({v+\xi\over2};{\i\pi\over2\epsilon}\right)\theta_2\left({v-\xi\over2};{\i\pi\over2\epsilon}\right)
  \over\theta_2\left({v+\xi\over2};{\i\pi\over2\epsilon}\right)\theta_1\left({v-\xi\over2};{\i\pi\over2\epsilon}\right)}
\label{S-matrix}
$$
coincide~\cite{Lukyanov:1995gs,Pugai:2002sft} with the $S$ matrices found in~\cite{Bazhanov:1990wh} for the neutral particles. The Bethe Ansatz derivation of these $S$ matrices in the context of the XYZ model was performed in~\cite{Fioravanti:2006sd}.

In the case of the $\RSOS(2,2s+1)$ models, i.e.
\eq$$
\xi={2\over2s-1},
\label{xi-spec}
$$
there are just two vacuums with $k=1$ ($l_\infty=s$) and $\nu=0,1$, and the subscript $k$ at the eigenvector can be omitted as it is done in~(\ref{matel-id}). In this case it is easy to check that $\sign(c_n^2)=(-1)^{\min(n,n^*)}$ (recall that $n^*=2s-1-n$) for $1\le n\le 2s-2$. Let us choose $c_n$ as follows:
\eq$$
c_n=|c_n|\times\Cases{\i^n,&n\le s-1;\\\i^{n-1},&s\le n\le2s-2.}
\label{cn-root}
$$
The main goal of the paper is to prove the identity
\eq$$
\bbT^{(n)}(u)\bigl|_{{\mathcal H}_{1l}}=\bbT^{(n^*)}(u)\bigl|_{{\mathcal H}_{1l}}
\label{Main-eq}$$
on the irreducible representations of the DVA, which is equivalent to~(\ref{matel-id}) in the RSOS model. First of all, it provides one more consistency check of the vertex operator approach to the RSOS lattice models. Second, the method of the proof may be useful in finding other identities between form factors.

Our proof is based on the free field representation for the DVA. In this representation the DVA is immersed into an appropriate Heisenberg algebra, while its irreducible modules $\cH_{kl}$ are realized by the Felder cohomologies of Fock spaces. The difficulty is related to the fact that $T^{(n)}$ and $T^{(n^*)}$ act differently on the Fock spaces, and their explicit free field representations exhibit no similarity. Recently, we obtained\cite{Lashkevich:2014qna} a similar kind of identity withing the algebraic approach of~\cite{Feigin:2008hs} to form factors of sine\-/Gordon theory. Nevertheless, the Felder resolution was not relevant in that approach, and the corresponding identities were valid subject to a complicated set of conditions. Finding a simpler and more transparent elliptic analog to those identities was one of the motivations of the present work.

In~\cite{Feigin:2007arXiv0705.0427} a family of generalized deformed Virasoro algebras was proposed. Define the current
\eq$$
\cT(z)=T(z)B_\sigma(z),
\label{cT-TB}
$$
where the current $B_\sigma(z)$, which depends on a parameter $\sigma$, is an exponential of the extra free boson with special commutation relations
\eq$$
B(z)=\lcolon\exp\sum_{m\ne0}{[m(\xi+1)]\alpha'_m z^{-m}\over m}\rcolon,
\qquad
[\alpha'_m,\alpha'_n]=m{[m\xi]\over[m(\xi+1)]}{[(\sigma-2)m]\over[\sigma m]}{[m]\over [2m]}\delta_{m+n,0}.
\label{B-def}
$$
For each $\sigma$ the current $\cT(z)$ together with one more current $\cT_2(z)$ generates an algebra with quadratic relations, which possesses nice properties from the purely algebraic point of view. Though most of the results below can be formulated in terms of the DVA itself, we will use this modified algebra because of several reasons. First, for $\sigma=1$ it simplifies the definition of fused currents $\cT^{(n)}$ corresponding to~$T^{(n)}$. Second, in contrast to the DVA (\ref{DVA-def}), it admits the limit $x\to\e^{-\i\pi/2}$, which corresponds to the construction of~\cite{Lashkevich:2014qna}. And the last, we expect that in this form the construction can be generalized to more complicated cases due to the relation of the modified algebra to the Ding\--Iohara algebra~\cite{Ding:1996mq}.

\section{Free field realization}
\label{sec-freefield}

The main tool of studying the deformed Virasoro algebra and the algebra defined in~\cite{Feigin:2007arXiv0705.0427} is the free field realization. Let us give its short review.

\subsection{Heisenberg algebra and Fock spaces}

Consider the Heisenberg algebra generated by a couple of oscillators $\beta^i_m$ ($i=0,1$; $m\in\Z\setminus\{0\}$) and zero modes $\cP$, $\cQ$ with the following commutation relations
\eq$$
\Gathered{
[\beta^i_m,\beta^i_n]=0
\qquad
[\beta^0_m,\beta^1_n]=-m{[m\xi]\over[m(\xi+1)]}x^m\delta_{m+n,0},
\\
[\beta^i_m,\cP]=[\beta^i_m,\cQ]=0,
\qquad
[\cP,\cQ]=-\i.
}\label{Halg-def}
$$
Define the Fock spaces $\hat\cF_P$ generated by the elements $\beta^i_{-m}$ ($m>0$) from the vacuum vectors~$|P\rangle$ such that
\eq$$
\beta^i_m|P\rangle=0\quad(m>0),
\qquad
\cP|P\rangle=P|P\rangle.
\label{bosvac-def}
$$
The conjugate vacuums
\eq$$
\langle P|\beta^i_{-m}=0\quad(m>0),
\qquad
\langle P|\cP=\langle P|P
\label{bosvac-conj-def}
$$
generate by the action of $\beta^i_m$ ($m>0$) the Fock spaces
\eq$$
\hat\cF^*_P\cong\hat\cF_{-P}.
\label{cFstar-cF}
$$

We will be interested in vacuums with special values of the `momentum'~$P$:
\eq$$
|k,l\rangle\equiv|P_{kl}\rangle,
\qquad
P_{kl}=(\xi+1)k-\xi l,
\qquad
k,l\in\Z.
\label{bosvac-kl-def}
$$
The corresponding Fock spaces will be called~$\hat\cF_{kl}$. For rational values of $\xi$ defined in~(\ref{rx-RSOS}) we have
\eq$$
\hat\cF_{k+p,l+q}=\hat\cF_{kl}.
\label{cF-period}
$$

In what follows it will be useful to represent the Heisenberg algebra defined above in terms of another set of oscillators
\eq$$
\alpha_m={[m]\over[2m]}(\beta^0_m-\beta^1_m),
\qquad
\alpha'_m={[m]\over[2m]}(x^m\beta^1_m+x^{-m}\beta^0_m),
\label{alpha-def}
$$
which satisfy the commutation relations
\eq$$
[\alpha_m,\alpha_n]=-[\alpha'_m,\alpha'_n]=m{[m\xi]\over[m(\xi+1)]}{[m]\over[2m]}\delta_{m+n,0},
\qquad
[\alpha_m,\alpha'_n]=0.
\label{alpha-commut}
$$
Since the oscillators $\alpha_m$ and $\alpha_m'$ mutually commute, the Fock spaces can be decomposed into tensor products of two their subspaces
\eq$$
\hat\cF_P\cong\cF_P\otimes\cF'.
\label{cFP-factor}
$$
Here by $\cF_P$ we denoted the space which is generated from the highest weight vectors $|P\rangle$ by the elements $\alpha_{-m}$ with $m>0$. Respectively, the subspace $\cF'$ is generated by the operators $\alpha'_{-m}$. For generic values of $x$ the first space provides a representation of the deformed Virasoro algebra, while the second one contains the factor $B(z)$ of~(\ref{cT-TB}).

\subsection{Current algebras}

The modified deformed Virasoro algebra is a subalgebra of the Heisenberg algebra~(\ref{Halg-def}) generated by the current~$\cT$ defined as
\eq$$\Aligned{
\cT(z)=\Lambda_0(z)+\Lambda_1(z).
}
\label{cT-def}
$$
This current is essentially a sum of the exponents of the free bosons introduced in the previous section
\eq$$
\Lambda_i(z)=x^{(1-2i)\cP}\lcolon\exp\sum_{m\ne0}{[m(\xi+1)]\beta^i_mz^{-m}\over m}\rcolon,
\quad i=0,1.
\label{Lambda01-def}
$$
Products of these operators can be reduced to normal products by using the formula
\eq$$
\Lambda_i(z')\Lambda_j(z)
=\lcolon\Lambda_i(z')\Lambda_j(z)\rcolon f_{ij}\left(z\over z'\right),
\label{LambdaLambda-prod}
$$
where
\eq$$
f_{00}(z)=f_{11}(z)=1,
\qquad
f_{01}(z)=f_{10}(z^{-1})={(1-x^{2\xi+2}z)(1-x^{-2\xi}z)\over(1-z)(1-x^2z)}.
\label{fpmpm-def}
$$
Note, that in the limit $x\to\e^{-i \pi/2}$ this give the construction, which appeared in the problem of finding form factors of local operators in the sine\-/Gordon model\cite{Lashkevich:2014qna}. In this sense we are studying the elliptic version of the dynamical symmetry algebra of the corresponding scaling model.

By using~(\ref{alpha-def}) the current $\cT(z)$ can be factorized into the product $T(z)B(z)$ in consistency with~(\ref{cT-TB}), thus defining the DVA generator~$T(z)$.

\subsection{Irreducible representations and Felder complex}

The Fock spaces $\cF_P$ are representations of the deformed Virasoro algebra. For generic values of $P$ they are irreducible, but for special values $P=P_{kl}$ they are not. The irreducible representations can be constructed as cohomologies of a special complex. The differential of this complex commutes with the DVA generators~$T(z)$. Let us briefly recall the construction.

First, consider the currents as~\cite{Lukyanov:1994re}
\eq$$
\Aligned{
E(z)
&=\e^{-2\i(\xi+1)\cQ}z^{-\cP+\xi+1\over\xi}
\lcolon\exp\sum_{m\ne0}{[m(\xi+1)]\over[m\xi]}{\beta^0_m-\beta^1_m\over m}z^{-m}\rcolon,
\\
F(z)
&=\e^{2\i\xi\cQ}z^{\cP+\xi\over\xi+1}
\lcolon\exp\sum_{m\ne0}{\beta^1_m-\beta^0_m\over m}z^{-m}\rcolon.
}\label{FE-def}
$$
In terms of these currents define the screening operators~\cite{Lukyanov:1996qs}
\eq$$
\Aligned{
X(z_0)
&=\oint{dz\over2\pi\i z}E(z){\lbbrack v-v_0-{1\over2}+\cP\rbbrack'\over\lbbrack v-v_0+{1\over2}\rbbrack'},
\\
Y(z_0)
&=\oint{dz\over2\pi\i z}F(z){\lbbrack v-v_0+{1\over2}-\cP\rbbrack\over\lbbrack v-v_0-{1\over2}\rbbrack},
\qquad
z=x^{2v},
\quad
z_0=x^{2v_0}.
}\label{screenings-def}
$$
In fact, the screening operators are only expressed in terms of the zero modes and oscillators $\alpha_m$, so that they act on the Fock spaces~$\cF_P$. For special values of momentum $P_{kl}$ they act as
\eq$$
X(z_0):\ \cF_{kl}\to\cF_{k-2,l},
\qquad
Y(z_0):\ \cF_{kl}\to\cF_{k,l-2}.
\label{XY-cF-action}
$$
It is known~\cite{Jimbo:1996vu} that on the modules $\cF_{kl}$ ($k\in\Z\setminus p\Z$, $l\in\Z\setminus q\Z$) the operators $X^{\bar k}(z_0)$ and $Y^{\bar l}(z_0)$ are nonzero and $z_0$\-/independent, if
$$
\Gathered{
1\le\bar k\le p-1,
\qquad
k-\bar k\in p\Z;
\\
1\le\bar l\le q-1,
\qquad
l-\bar l\in q\Z.
}
$$
These operators commute with the DVA generators:
\eq$$
[X^{\bar k},T(z)]|_{\cF_{kl}}=0,
\qquad
[Y^{\bar l},T(z)]|_{\cF_{kl}}=0.
\label{XYT-commut}
$$
Besides,
\eq$$
X^{p-\bar k}X^{\bar k}|_{\cF_{kl}}=0,
\qquad
Y^{q-\bar l}Y^{\bar l}|_{\cF_{kl}}=0,
\label{XX-YY-zero}
$$
so that these operators can be considered as differentials that act on a special complex. Take, for example, the action of powers of the $X(z_0)$ operator on the Fock spaces. Let $1\le k\le p-1$, $1\le l\le q-1$. Then the sequence
$$
\cdots\xleftarrow{\quad X^k\quad}\cF_{k-2p,l}\xleftarrow{\quad X^{p-k}\quad}\cF_{-k,l}
\xleftarrow{\quad X^k\quad}\cF_{kl}\xleftarrow{\quad X^{p-k}\quad}\cF_{2p-k,l}\xleftarrow{\quad X^k\quad}\cdots
$$
form a complex, called the Felder complex. Just as in the case of the usual Virasoro algebra~\cite{Felder:1988zp}, the cohomologies of this complex vanish except for one term
\eq$$
\cH_{kl}=\Ker_{\cF_{kl}}X^k/\Im_{\cF_{2p-k,l}}X^{p-k}.
\label{cHkl-def}
$$
The space $\cH_{kl}$ is an irreducible representation of the DVA~\cite{Pugai:2004mi}. The traces $\Tr_{\cH_{kl}}$, used in the previous section, are defined in terms of traces over Fock spaces:
\eq$$
\Tr_{\cH_{kl}}x^{4H}\Phi=\sum_{j\in\Z}\left(\Tr_{\cF_{k+2pj,l}}x^{4H}\Phi-\Tr_{\cF_{-k+2pj,l}}x^{4H}\Phi\right)
\label{TrcHkl-def}
$$
for any admissible product $\Phi:\cF_P\to\cF_P$ of the operators $\Phi(u)^{l'}_l$ and $\bbT^{(n)}(u)$ realized in terms of the oscillators~$\alpha_m$ and zero modes $\cP$,~$\cQ$.

The cohomologies are trivially extended to the complex of the spaces~$\hat\cF_{kl}$:
\eq$$
\hat\cH_{kl}=\Ker_{\hat\cF_{kl}}X^k/\Im_{\hat\cF_{2p-k,l}}X^{p-k}=\cH_{kl}\otimes\cF'.
\label{hatcHkl-def}
$$
This extends the construction to the algebra of the currents~$\cT(z)$, which will be used below.

In the case $p=2$, $q=2s+1$, which is our main subject, we are interested in odd values of~$k$. Thus, the operator
\eq$$
X=(-1)^{l-1+{k-1\over2}}\int{dz\over2\pi\i z}\,E(z),
\qquad
k\text{ odd},
\label{X-ribbon}
$$
is $z_0$\-/independent, commutes with the DVA and satisfies the equation $X^2=0$. The Felder complex looks like
$$
\cdots\xleftarrow{\quad X\quad}\cF_{-3,l}\xleftarrow{\quad X\quad}\cF_{-1,l}
\xleftarrow{\quad X\quad}\cF_{1l}\xleftarrow{\quad X\quad}\cF_{3l}\xleftarrow{\quad X\quad}\cdots
$$
and has the only nonzero cohomology
$$
\cH_{1l}=\Ker_{\cF_{1l}}X/\Im_{\cF_{3l}}X,
\qquad
1\le l\le 2s.
$$

\subsection{Fused currents}

We are interested in studying the fusion currents, which in the corresponding scattering theory correspond to bound states.
\eq$$
\cT^{(n)}(z)
=\prod^{\substack{\curvearrowright\\ n-1}}_{j=0}\cT(x^{(n-1-2j)\xi}z).
\label{cT(n)-def}
$$
These currents are related with the $T^{(n)}(z)$ currents defined earlier as follows. Let
\eq$$
B^{(n)}(z)=\lcolon\prod^{n-1}_{j=0}B(x^{(n-1-2j)\xi}z)\rcolon.
\label{B(n)-def}
$$
Then
\eq$$
\cT^{(n)}(z)=T^{(n)}(z)B^{(n)}(z).
\label{cT(n)-T(n)-rel}
$$
Explicitly, the currents $\cT^{(n)}(z)$ read
\eq$$
\cT^{(n)}(z)
=\sum^n_{j=0}f^{(n)}_j\Lambda^{(n)}_j(z),
\label{T(nu)-def}
$$
where
\Align$$
\Lambda^{(n)}_j(z)
&=\lcolon\prod^{j-1}_{i=0}\Lambda_1(x^{-(n-1-2i)\xi}z)
\prod^{n-1}_{i=j}\Lambda_0(x^{-(n-1-2i)\xi}z)\rcolon
\notag
\\
&=x^{(n-2j)\cP}\lcolon\exp\sum_{m\ne0}{[m(\xi+1)]\over[m\xi]}\left(x^{m(n-j)\xi}[mj\xi]\beta^1_m
+x^{-mj\xi}[m(n-j)\xi]\beta^0_m\right){z^{-m}\over m}\rcolon.
\label{Lambda(nu)-def}
$$
Obviously, $\Lambda^{(1)}_j(z)=\Lambda_j(z)$ for $j=0,1$. The numeric coefficients $f^{(n)}_j$ in the equation (\ref{T(nu)-def}) can be easily derived from the equations (\ref{LambdaLambda-prod}),~(\ref{fpmpm-def}). Explicitly, they read
\eq$$
f^{(n)}_j=\prod^{j-1}_{k=0}{[(n-k)\xi][k\xi-1]\over[(k+1)\xi][(n-k-1)\xi-1]}.
\label{f(nu)-def}
$$
It is easy to check that
$$
f^{(n)}_j=f^{(n)}_{n-j},
\qquad
f^{(n)}_0=f^{(n)}_n=1.
$$
For a generic value of the parameter $\xi$ there are infinitely many independent fused currents of the form~(\ref{T(nu)-def}). On the contrary, when $\xi$ is a positive rational number we expect reduction formulas, which relate different currents. In the next section we prove that in the case~(\ref{xi-spec}) the reduction formula~(\ref{Main-eq}) holds.

\section{The homotopy operator}
\label{sec-tau-operator}

The free field expressions~(\ref{T(nu)-def}) for fused currents $\cT^{(n)}$ and $\cT^{(n^*)}$ look essentially different. Here we will show that there exists an alternative construction for the fusion currents where different $\cT^{(n)}$ are represented uniformly.

Let us define exponential operators of a special form
\eq$$
\tau_n(z)
=\e^{2\i(\xi+1)\cQ}z^{\cP+\xi+1\over\xi}
\lcolon\exp\sum_{m\ne0}{[m(\xi+1)]\over[m\xi]}{\left(x^{mn\xi}\beta^1_m-x^{-mn\xi}\beta^0_m\right)}{z^{-m}\over m}\rcolon.
\label{tau-def}
$$
Note that for $n=0$ the corresponding exponent is the inverse screening operator,
$$
\tau_0(z)=z^{2(\xi+1)/\xi}\lcolon E^{-1}(z)\rcolon,
$$
as it is clear from the definition (\ref{FE-def}). They are characterized by the property that the corresponding operator product expansion with the screening current $E(z)$ are of a very nice form
\eq$$
\Aligned{
\tau_n(z')E(z)
&=z^{\prime-{2(\xi+1)\over\xi}}g_n\left(z\over z'\right)h_n\left(z\over z'\right)\lcolon\tau_n(z')E(z)\rcolon,
\\
E(z)\tau_n(z')
&=z^{-{2(\xi+1)\over\xi}}g_n\left(z'\over z\right)h_n\left(z'\over z\right)\lcolon\tau_n(z')E(z)\rcolon,
}\label{taunE-op}
$$
where
\eq$$
g_n(z)={(x^{2-(n-2)\xi}z;x^{2\xi})_\infty\over(x^{n\xi-2}z;x^{2\xi})_\infty},
\qquad
h_n(z)=\prod^n_{j=0}{1\over 1-x^{(n-2j)\xi}z}.
\label{gnhn-def}
$$
There are infinitely many poles in the right hand sides of~(\ref{taunE-op}) due to the function~$g_n(z)$. Let us, however, concentrate attention to the poles at the points $z=z'x^{-(n-2j)\xi}$. The remarkable fact is that their residues are proportional to exponential operators $\Lambda^{(n)}_j(z)$, which enter the fused currents~$\cT^{(n)}$:
\eq$$
\lcolon\tau_n(z)E(zx^{-(n-2j)\xi})\rcolon=z^{2{(\xi+1)/\xi}}x^{-(\xi+1)(n-2j)}\Lambda^{(n)}_j(z).
\label{taunE-res}
$$
By gathering all the results we arrive to the following integral representation for the currents:
\eq$$
\cT^{(n)}(z) =K_n^{-1}\oint_{\cC_n}{dw\over2\pi\i w}\,\tau_n(z)E(w),
\label{taunE-T}
$$
where the contour $\cC_n$ encloses all poles $w=zx^{-(n-2j)\xi}$ ($j=0,1,\ldots,n$) and does not enclose any other poles, and
\eq$$
K_n=x^{-n(\xi+1)}g_n(x^{-n\xi})h_n(x^{-(n+1)\xi}).
\label{K(nu)-def}
$$
For generic values of the parameter $\xi$ the formula (\ref{taunE-T}) looks rather artificial because its contour encloses a finite subset of the infinite set of poles. But for special values~(\ref{xi-spec}) of $\xi$ the infinite `tail' of poles cancels out. Indeed, due to the identity
\eq$$
g_n(z)=h_{n^*}(z)
\quad\text{for}\quad\xi={2\over2s-1}
\label{taunE-ratio-ribbon}
$$
we have
\eq$$
\tau_n(z')E(z)=-E(z)\tau_n(z')
=\lcolon\tau_n(z')E(z)\rcolon
z^{\prime\,-2s-1}h_n\left(z\over z'\right)h_{n^*}\left(z\over z'\right).
\label{taunE-op-spec}
$$
Thus, these products have a finite set of poles. Notice, that they are just the poles enclosed by contour $\cC=\cC_n+\cC_{n^*}$. Let us integrate $\tau_n(z)E(w)$ with respect to $w$ along the contour~$\cC$. The contribution of the contour $\cC_n$ is given by~(\ref{taunE-T}). To obtain the contribution of the contour $\cC_{n^*}$, we need some arrangements.

Let us introduce the exponential operator $\mu_n$, which is the `ratio' of the operators $\tau_n$ and~$\tau_{n^*}$:
\eq$$
\tau_n(z)=\lcolon\mu_n(z)\tau_{n^*}(z)\rcolon,
\label{tau-mutau}
$$
Explicitly,
\eq$$
\mu_n(z)=\lcolon\exp\sum_{m\ne0}{[m(\xi+1)][m(\xi n-1)]\over[m\xi]}(x^m\beta^1_m+x^{-m}\beta^0_m)z^{-m}\rcolon.
\label{mu-def}
$$
An important property of $\mu_n(z)$ is that it is expressed in terms of $\alpha'_m$ and, hence, commutes with~$\alpha_m$:
$$
[\mu_n(z),\beta^0_m-\beta^1_m]=0,
$$
This means that the multiplication by $\mu_n(z)$ does not change the operator products with the screening currents. Thus, we obtain
\eq$$
\int_{\cC_{n^*}}{dw\over2\pi\i w}\,\tau_n(z)E(w)
=\int_{\cC_{n^*}}{dw\over2\pi\i w}\,\lcolon\mu_n(z)\tau_{n^*}(z)\rcolon E(w)
=K_{n^*}\lcolon\mu_n(z)\cT^{(n^*)}(z)\rcolon.
\label{taunE-muT}
$$
As a result, by integrating along the contour $\cC$ we get an expression that contains both $\cT^{(n)}$ and $\cT^{(n^*)}$ currents:
\eq$$
\oint_\cC{dw\over2\pi\i w}\,\tau_n(z)E(w)
=K_n\cT^{(n)}(z)+K_{n^*}\lcolon\mu_n(z)\cT^{(n^*)}(z)\rcolon.
\label{taunE-T+muT}
$$
Since the contour $\cC$ encloses all poles of the integrand operator product, it can be split into a difference $\cC^+-\cC^-$, so that the contours $\cC^\pm$ are counterclockwise and enclose zero. The contour $\cC^-$ does not enclose any pole of $\tau_n(z)E(w)$ while $\cC^+$ encloses all of them. From (\ref{X-ribbon}) we conclude that the l.h.s.\ of~(\ref{taunE-T+muT}) on the Fock spaces $\hat\cF_{kl}$ with odd $k$ is nothing but an commutator:
\eq$$
\left.(-1)^{l+{k-1\over2}}[\tau_n(z),X]\right|_{\hat\cF_{kl}}
=\left.K_n\cT^{(n)}(z)+K_{n^*}\lcolon\mu_n(z)\cT^{(n^*)}(z)\rcolon\right|_{\hat\cF_{kl}},
\qquad
k\text{ odd}.
\label{X-taun-commut}
$$
Now we can show that on the cohomologies the l.h.s.\ is zero. Indeed, let $|v\rangle\in\Ker_{\hat\cF_{kl}}X$. Then $[\tau_n(z),X]|v\rangle=-X\tau_n(z)|v\rangle$ is an exact vector. Hence, the l.h.s.\ of (\ref{X-taun-commut}) is zero on the cohomology~$\hat\cH_{kl}$. In terms of the homology theory the operator $\tau_n(z)$ is a \emph{homotopy} operator. Therefore, we have
\eq$$
\left.\cT^{(n)}(z)\right|_{\hat\cH_{kl}}=\left.C_n\lcolon\mu_n(z)\cT^{(n^*)}(z)\rcolon\right|_{\hat\cH_{kl}},
\qquad
k\text{ odd,}
\label{TT-Hkl-rel}
$$
where
\eq$$
C_n=-{K_{n^*}\over K_n}=-\prod^{n^*}_{j=n+1}{[1-j\xi]\over[j\xi]}.
\label{Cnu-def}
$$
Since the only nonzero cohomology corresponds to the case $k=1$, $1\le l\le 2s$, the identity is nontrivial for this case only.

In fact, the free boson $\alpha'_m$ completely decouples from the construction so that the identity (\ref{TT-Hkl-rel}) can be written in terms of the DVA currents~$T^{(n)}$:
\eq$$
\left.T^{(n)}(z)\right|_{\cH_{1l}}=\left.C_nT^{(n^*)}(z)\right|_{\cH_{1l}}.
\label{TT-DVA-Hkl-rel}
$$
It is straightforward to check that
\eq$$
C_n={c_{n^*}\over c_n},
\label{Cn-cn-rel}
$$
which completes the proof of eq.~(\ref{Main-eq}).

\section{The \texorpdfstring{$x\to\e^{-\i\pi/2}$}{x->exp(-i pi/2)} case: two limits}
\label{sec-fflimit}

The operators $\tau_n(z)$ were introduced in~\cite{Lashkevich:2014qna} in the framework of a free field construction for form factors of the sinh- and sine\-/Gordon models. This construction is based on the formal limit of the free field representation described in section~\ref{sec-freefield} as $x\to\e^{-\i\pi/2}$ for generic real values of~$\xi$. In this limit the DVA is not defined, but the algebra of~\cite{Feigin:2007arXiv0705.0427} still exists for $\sigma=1$. In this limiting construction the operator $\tau_n(z)$ again generates the bound state currents $\cT^{(n)}(z)$, but the r.h.s.\ of the expression (\ref{taunE-T}) takes the form proportional to the commutator $[\tau_n(z),X]$. The difference with (\ref{X-taun-commut}) is related to the order of limits. Namely, take the limit in two stages. Let $\varepsilon,\delta>0$ and
$$
x=\e^{-\i\pi-\varepsilon},
\qquad
\xi=\xi_0-\i\delta,
\qquad
\xi_0\in\R.
$$
First take the limit $\varepsilon\to0$, and then $\delta\to0$. This limit corresponds to the construction proposed in~\cite{Lashkevich:2014qna}. Let us call it `limit~A'. We can see that in this limit
\eq$$
g_n(z)\xrightarrow[\ve\to0]{}{(-x^{-(n-2)\xi}z;x^{2\xi})_\infty\over(-x^{n\xi}z;x^{2\xi})_\infty}
=h_{n-2}^{-1}(-z).
\label{taunE-ratio-lim}
$$
On the contrary, if we assume $\delta=0$, $\xi_0=2/(2s-1)$ for finite $\varepsilon$, we will just obtain~(\ref{taunE-ratio-ribbon}). Then we may take the limit~$\varepsilon\to0$. We will call it `limit~B'. The expressions (\ref{taunE-ratio-lim}) and (\ref{taunE-ratio-ribbon}) are incompatible, which means noncommutativity of the two limits.

In~\cite{Lashkevich:2014qna} special matrix elements were considered. Namely, consider the commutative algebra $\cA$, generated by an infinite set of elements~$a_{-m}$ with $m>0$. Consider the right action of this algebra on the Fock space $\cF_P$ and the left action on~$\cF^*_P$. The representatives of the right action will be denoted by the same letters $a_{-m}$ with a bar, while the elements of the right action by the letters $\bar a_{-m}$:
\eq$$
a_{-m}={\beta^0_m-\beta^1_m\over[m\xi]},
\qquad
\bar a_{-m}={\beta^0_{-m}-\beta^1_{-m}\over[m\xi]},
\qquad m>0,
\label{am-def}
$$
so that
\eq$$
\cF_P=\cA|P\rangle,
\qquad
\cF^*_P=\langle P|\cA.
\label{cA-action}
$$
For any element $h\in\cA$ define the vector $|h:P\rangle=\bar h|P\rangle\in\cF_P$ and, correspondingly, $\langle h:P|=\langle P|h\in\cF^*_P$. Then consider the matrix elements
\eq$$
J^{h\bar h'}_P(n_1,z_1;\ldots;n_N,z_N)=\langle h:P|\cT^{(n_1)}(z_1)\cdots\cT^{(n_N)}(z_N)|h':P\rangle.
\label{Jhh'-def}
$$
They are easily calculated for any given pair $h,h'$ by means of the commutation relations
\subeq{\label{am-commut}
\Gather$$
[a_{-m},\Lambda_i(z)]=(-)^ix^{(-)^{i+1}m}z^m\Lambda_i(z),
\qquad
[\Lambda_i(z),\bar a_{-m}]=(-)^ix^{(-)^im}z^{-m}\Lambda_i(z),
\label{am-Lambdai-commut}
\\
[a_{-m},\bar a_{-n}]={m[2m]\over[m][m(\xi+1)][m\xi]}\delta_{mn}.
\label{am-baran-commut}
$$}
These matrix elements have a finite limit as $|x|\to1$, and this limit is used in calculation of breather form factors of local operators in the sine\-/Gordon model. Omitting details, we may say that the set of functions $J^{h\bar h'}_P$ with arbitrary number of arguments uniquely defines a local operator, which will be denoted as $\cO^{h\bar h'}_P(x)$ (here $x$ is a point in the flat two\-/dimensional space\-/time). In the case~(\ref{xi-spec}) the sine\-/Gordon model admits a reduction to the so called massive $\Phi_{13}$ perturbation of the $M(2,2s+1)$ minimal conformal model\cite{Smirnov:1990vm}. An operator $\cO^{h\bar h'}_P$ is consistent with the reduction if%
\footnote{In~\cite{Lashkevich:2014qna} a more general class of operators that contains derivative in $p$ was considered, but here we limit ourselves by a narrower class.}
\eq$$
J^{h\bar h'}_P(n_1,z_1;n_2,z_2;\ldots)=C_{n_1}J^{h\bar h'}_P(n^*_1,z_1;n_2,z_2;\ldots).
\label{JJ-rel}
$$
Note that the coefficient $C_n$ has the same limiting value in the limits~A and~B. However, the functions $J^{h\bar h'}_P$ have different limits due to different limits of the commutation relation~(\ref{am-baran-commut}). Indeed, for odd integer $m/(2s-1)$ the commutator $[a_m,\bar a_m]$ vanishes in the limit~A and is nonzero in the limit~B:
\eq$$
[a_m,\bar a_m]=-{m^2\over4},
\qquad
m\in(2s-1)(2\Z+1)
\quad\text{(limit~B).}
\label{am-baran-oddbad}
$$
For even integer $m/(2s-1)$ such commutator diverges as $\varepsilon^{-2}$ in both limits.

The operator $\cO^{h\bar h'}_P$ can only be consistent with the reduction for the special values $P=P_{kl}$ with $k=1$, $1\le l\le2s$ or equivalent according to~(\ref{cF-period}) regions. It is a necessary but not sufficient condition. The sufficient condition in the limit~A was found in~\cite{Lashkevich:2014qna}, but it has a complicated form. Now we show that in the limit~B the condition is simpler. Namely, from (\ref{X-taun-commut}) it immediately follows that the operator $\cO^{h\bar h'}_P$ is consistent with the reduction, if
\eq$$
|h:-k,-l\rangle\in\Ker_{\cF_{-k,-l}}X,
\qquad
|h':k,l\rangle\in\Ker_{\cF_{kl}}X.
\label{hh'-condition}
$$
Note that these kernel spaces do not contain vectors generated by the elements $\bar a_{-m}$ with $m=(2s-1)k$ and $k\in2\Z$. Indeed, in the limit~B the commutator of $X$ with these elements diverge as $\varepsilon^{-1}$, so that they can enter with the coefficient of the order~$\varepsilon$. On the contrary, the elements with $k\in2\Z+1$ may generate vectors in these spaces, but they do not correspond to commuting integrals of motion as $\bar a_m$ elements with other odd values of $m$ due to nonzero limit of the commutator~(\ref{am-baran-commut}) in this case.

\section{Conclusion}
\label{sec-conclusion}

The free field representation for the deformed Virasoro algebra was studied and the coincidence of the operators $\bbT^{(n)}(z)$ and $\bbT^{(n^*)}(z)$ ($n^*=2s-1-n$) on the cohomologies of the Felder resolution was proved by constructing a homotopy operator $\tau^{(n)}(z)$. As a result the traces (\ref{E-tr-rel}), which represent form factors in the lattice $\RSOS(2,2s-1)$ models, are unchanged under the substitution $n_i\to n^*_i$ for any~$i$.

Besides, the limiting case $x\to\e^{-\i\pi/2}$, which describes the sinh\-/Gordon model, was considered. It turned out that the result essentially depends on the order of limits $x\to\e^{-\i\pi/2}$ and $\xi\to2/(2s-1)$. If the former limit is taken first, the construction of~\cite{Lashkevich:2014qna} is obtained. The restriction imposed by the reduction condition is rather complicated. If, on the contrary, we first specialize to the rational values of $\xi$ and then take the limit for $x$, the reduction condition takes a simple form of restricting to the kernel of the screening operator.

We expect that this approach can be extended to a larger set of lattice and field theory models, where the closeness of system of particles should be proved in the free field representation. The first example to be treated should be the regime~II unitary RSOS models~\cite{Jimbo:2000ff} and their field theory counterparts, the scaling $\Z_N$ symmetric Ising models~\cite{Lashkevich:2018jmo}.

\section*{Acknowledgments}

The work was supported by Russian Foundation of Basic Research under the grant 18--02--01131. The work of M.~L.\ and Y.~P.\ was supported, in part, by grants of the Simons Foundation. The work of J.~S.\ was also supported in part by the grant 19K03512 from the Japan Society for the Promotion of Science (JSPS), while the work of Y.~T.\ was supported by the grant KAKENHI 26800064 from the JSPS.


\end{document}